# Bimodal drop size distributions during the early stages of shear induced coalescence

Verena E. Ziegler, Bernhard A. Wolf*

Institut für Physikalische Chemie der Johannes Gutenberg-Universität,
Jakob-Welder-Weg 13, D-55099 Mainz, Germany
FAX: +(0)6131-39-24640
Electronic mail: bernhard.wolf@uni-mainz.de

**Abstract:** Drop sizes and drop size distributions were determined by means of an optical shear cell in combination with an optical microscope for the systems polyisobutylene/poly(dimethylsiloxane) [I] and poly(dimethyl-*co*-methylphenylsiloxane)/poly(dimethylsiloxane) [II] at low concentrations of the suspended phases and at different constant shear rates ranging from 10 to 0.5 $s^{-1}$. After pre-shearing the two-phase mixtures [I: 50 $s^{-1}$; II: 100 $s^{-1}$] for the purpose of producing small drop radii, the shear rate was abruptly reduced to the preselected value and coalescence was studied as a function of time. In all cases one approaches dead end drop radii, i.e. breakup is absent. The drop size distributions are for sufficiently long shearing always unimodal, but within the early stages of coalescence they are in some cases bimodal; the shape of the different peaks is invariably Gaussian. The results are discussed by means of Elmendorp diagrams and interpreted in terms of collision frequencies and collision efficiencies.

## 1. Introduction

The temporal development of the size of drops and their size distribution in mechanical fields is of central importance for every production process involving two phase systems. Concerning the final status resulting for steady shear there exist two conceptually clearly distinguishable cases: The establishment of stationary states (the rate of coalescence equals the rate of breakup) and the approach of ultimate size exclusively via coalescence (because of the lack of breakup processes). The radii resulting in the former case are subsequently called steady state radii, in contrast to that we speak of dead end drop radii in the latter case (in contrast to the term pseudo steady drop radii sometimes used in the literature).

Research on the effect of shear on the breakup of drops has a long-standing tradition. It started with the work of Taylor [1] and primarily dealt with the disruption of single drops. Studies along these lines was for example continued by Rumscheidt and Mason [2] and Torza et al. [3]. DeBruijn [4] reported an empirical relation (based on experimental data of his own and of Grace [5]) for the dependence of the critical drop size on shear rate. This dependence describes at which radius a drop becomes susceptible to breakup at a given shear rate.

Despite early research on coalescence [6] a more detailed investigation of this phenomenon commenced considerably later. An important step to a better understanding was the establishment of the model of film drainage by Chesters [7,8], adopted to polymers by Janssen [9]. It was above all the coalescence of blends containing only small volume fractions of the drop phase under simple shear flow, which was studied in great detail. For the present work the results of the following authors were particularly relevant. Grizzuti and Bifulco [10], Minale et al. [11,12], Vinckier et al. [13], Rusu and Peuvrel-Disdier [14], Ramic et al. [15] and Lyu and Bates [16,17].

Some groups [13,14,16,17] observed only coalescence in their experiments. According to their interpretation one reaches a dead end drop size because of vanishing coalescence efficiency. In contrast to these authors Ramic et al. [15] stated that the drop will in the course of coalescence inevitably become large enough to undergo breakup for sufficiently large measuring times (in the reported experiments up to 100 000 strain units). This means that stationary state radii are finally established. Grizzuti et al. [10] and Minale and co-workers [11,12] have reported that both types of behavior can be observed with a given system, depending on the particular experimental conditions (volume fractions of the drop phase and shear rate).

The present contribution deals above all with the question, how the size and the distribution of the drops of the minor phase change with time via coalescence in the absence of breakup processes. For this study we have chosen a blend of two homopolymers and a mixture of a homopolymer and a copolymer. In both cases the viscosities of the components where chosen such that the coalescence processes can be well followed by means of a light microscope and that the drops do not break within the experimentally accessible range of shear rates.

## 2. Theoretical Background

The variables which influence the number and shape of drops of a liquid suspended in another liquid when the system is sheared are the interfacial tension $\sigma$, the matrix viscosity $\eta_m$, the viscosity of the drop $\eta_d$, the radius $R_0$ of the drop in quiescent state and the shear rate $\dot{\gamma}$. The capillary number $Ca$ gives the ratio between the viscous force, which tends to deform and to break the drop, and the counteracting interfacial force, which tries to minimize the surface of the drop. The capillary number is defined as

$$Ca = \frac{\dot{\gamma} \eta_m R_0}{\sigma} \quad (1)$$

***Breakup:*** As the capillary number surpasses a critical value $Ca_{crit}$, the viscous force overcomes the interfacial force and the drop gets unstable against breakup. According to Taylor [1,18] $Ca_{crit}$ is given by

$$Ca_{crit} = 0.5 \frac{16\lambda + 16}{19\lambda + 16} \quad (2)$$

where

$$\lambda = \frac{\eta_d}{\eta_m} \quad (3)$$

DeBruijn [4] fitted experimental data of Grace [5] and of his own and obtained the following equation for $Ca_{crit}$

$$\begin{aligned}\log Ca_{crit} = \\ -0{,}506 - 0{,}0994 \log \lambda \\ + 0{,}124 (\log \lambda)^2 - \frac{0{,}115}{\log \lambda - \log 4.08}\end{aligned} \quad (4)$$

If a reliable value for $Ca_{crit}$ is accessible via the viscosity ratios $\lambda$ (eqs (3) and (4)) of a given system and $\sigma$ is known, the radius of the drops that are just stable against breakup at a given shear rate can be easily forecast by means of eq (1). This radius is in the following called $R^{\text{DeBruijn}}$.

*Coalescence:* The process of shear induced coalescence can be described by the equations of Janssen [9] following the models of Chester [7,8,19]. In order to coalesce two drops have firstly to meet and secondly stay long enough together to allow the separating matrix film to drain; for sufficiently long measuring times the collision probability can be set 1. The collision time is inversely proportional to the shear rate; during this period the matrix film has to drain down to a critical thickness $h_{\text{crit}}$ at which it becomes unstable due to van der Waals-forces and disrupts. If the collision time is too short or if the drops are too big, the amount of matrix film, which has to be removed becomes too large and the film thickness will not reach the critical value, i.e. the two drops come apart without any change. Three models for the maximum size the drop can reach via coalescence are distinguished regarding the mobility of the interface: immobile interface (IMI), partially mobile interface (PMI) and fully mobile interface (FMI):

IMI:
$$R = \left(\frac{8}{9}\right)^{1/4} h_{krit}^{1/2} \left(\frac{\eta_m \dot\gamma}{\sigma_{12}}\right)^{-1/2} \qquad (5)$$

PMI:
$$R = \left(\frac{16}{3}\right)^{1/5} \left(\frac{h_{krit}}{\lambda}\right)^{2/5} \left(\frac{\eta_m \dot\gamma}{\sigma_{12}}\right)^{-3/5} \qquad (6)$$

FMI:
$$R \ln\left(\frac{R}{h_{krit}}\right) = \left(\frac{2}{3}\right) \left(\frac{\eta_m \dot\gamma}{\sigma_{12}}\right)^{-1} \qquad (7)$$

For binary polymer blends the PMI-model gives the best accordance with the experimental data in the most cases ( [10-12,14]) but the IMI-model can also be suited [20,21]. The critical film thickness and therefore the drop size should be independent of concentration but this does not necessarily hold true as shown by results of Sundararaj and Macosko [22], Minale et al. [11] and Vinckier et al. [13].

## 3. Experimental Part

Polyisobutylene (PIB 3) was kindly donated from BASF, Germany, the two poly(dimethylsiloxane)s (PDMS 152 and PDMS 48) from Wacker, Germany. The poly(dimethyl-co-methylphenylsiloxane) (COP 26*) was purchased from Roth, Germany. The numbers after the abbreviation state the weight average molar mass in kg/mol. These data were obtained by means of GPC measurements in THF (PIB) or toluene (PDMS and COP 26*), using polystyrene standards and applying the universal calibration. Because the Kuhn-Mark-Houwink parameters were not known for COP 26*, we used the corresponding data for PDMS to obtain an apparent molar mass indicated by an asterisk. According to [1]H-NMR measurements this polymer contains 30% methylphenyl units. At room temperature COP 26* is practically immiscible with PDMS 48. All experiments were done at 25 °C. Viscosities were measured with the shear controlled rheometer CV 100 from Haake, Germany, and with the stress controlled rheometer AR 1000 from TA instruments, USA. Within the investigated range of shear rates (0 – 50 s$^{-1}$ for PIB 3 / PDMS 152 and 0 – 100 s$^{-1}$ for COP 26* / PDMS 48) all polymers behave Newtonian. The characteristic data of the polymers are collected in Tab. 1.

**Table 1**: Molar masses (as obtained from GPC measurements) and viscosities of the polymers

| polymer | $M_n$ / kg mol$^{-1}$ | $M_w$ / kg mol$^{-1}$ | $\eta^{25°C}$ / Pa s |
|---|---|---|---|
| PDMS 152 | 80.9 | 152.1 | 69.7 |
| PIB 3 | 1.3 | 2.6 | 26.9 |
| PDMS 48 | 29.5 | 48.5 | 2.2 |
| COP 26* | 5.2* | 26.4* | 1.2 |

The interfacial tensions $\sigma$ at 25 C° were determined by the method of drop retraction as described by Guido and Villone [23] using the optical shear cell CSS450 from Linkam Scientific, UK. Further details can be found in the literature [24]. For PDMS 152 drops (5 vol.-%) in PIB 3 we obtained 2.25 mNm$^{-1}$, 5 Vol-% COP 26* in PDMS 48 show $\sigma = 0.49$ mNm$^{-1}$ [24].

The morphology development during shear was observed with the optical shear cell CSS 450. The samples were prepared by stirring the blends with a spatula by hand for 3 minutes. After that vacuum was applied until all air bubbles were removed; the time required for that purpose ranged from 5 min up to 4 hours. Thereafter the sample was placed into the shear cell and was pre-sheared for 5 min at 100 (COP 26*/PDMS 48) or at 50 s$^{-1}$ (PIB 3/PDMS 152). This procedure suffices to eradicate the sample history. After that treatment $\dot{\gamma}$ was rapidly reduced to a preselected value and the morphology development was observed by taking digital images of the sheared sample at different times.

These images were analyzed with the help of the software Optimas 6.1, Mediacybernetics, USA, by extracting manually the main axis $L_{app}$ of at least 200 drops. With the CSS 450 instrument the direction of observation is perpendicular to the plane of shear. Because the main axis of a deformed drop is normally not aligned in the plane of shear, it is with this setup only possible to see the projection of the main axis. For that reason we have established a correlation function between $L_{app}$ and $R_0$ in the following manner. The blend was sheared at a certain shear rate until the drops took their equilibrium shape. Then shear was stopped and images were taken very rapidly during the relaxation period. From the last image of the sheared drop and an image taken after its complete relaxation the elongated axis and the diameter in quiescent state were evaluated for about 50 drops. In this manner one obtains a correlation between these two parameters for the different shear rates, which enables the calculation of $R_0$ from the observed $L_{app}$.

We used two different averages of drop size to quantify polydispersity: The number average $R_N$ and the volume average $R_V$; furthermore we determined the mean value of the Gaussian fit to the drop size distribution $R_G$. The following relations hold true

$$R_N = \frac{\sum_{i=1}^{\infty} n_i R_i}{\sum_{i=1}^{\infty} n_i} \qquad (8)$$

$$R_V = \frac{\sum_{i=1}^{\infty} n_i R_i^4}{\sum_{i=1}^{\infty} n_i R_i^3} \qquad (9)$$

$$y = y_0 + \frac{A}{SD\sqrt{2\pi}} e^{-\frac{(R-R_G)^2}{2SD^2}} \qquad (10)$$

$SD$: standard deviation

## 4. Results and Discussion

The following results refer to the blends PIB 3/PDMS 152 with volume fraction $\varphi_{PDMS\ 152} = 0.050$ and COP 26*/PDMS 48 with $\varphi_{PDMS\ 48} = 0.051$. In both cases the viscosity ratio is larger than unity. The investigated shear rates after step down from 50/100 s$^{-1}$ are 1, 2, 5 and 10 s$^{-1}$ for PIB 3/PDMS 152 and 0.5, 1, 2, 5, 10 and 20 s$^{-1}$ for COP 26*/PDMS 48. This section is divided into three parts, the first deals with drop sized and drop size distributions, the second discusses the time development of the mean radii and the third concerns the shear rate dependence of the dead end radii.

### 4.1. Drop size distribution and bimodality

An example for the images taken during the course of the experiments are shown in Fig. 1 together with the obtained drop size distributions for $\varphi_{PDMS\ 152} = 0.050$ after a step down of $\dot{\gamma}$ from 50 to 5 s$^{-1}$. From the very beginning one observes a very narrow drop size distribution, which shifts to higher radii with progressing time. It can be fitted easily by the equation of Gauss (eq (10)).

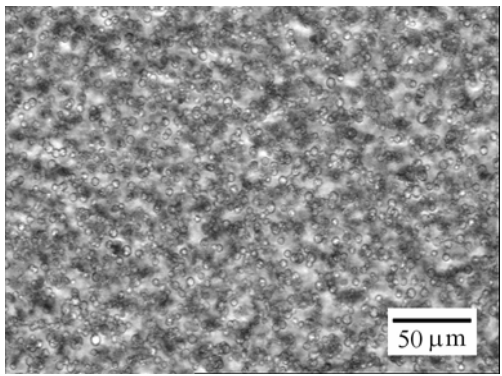
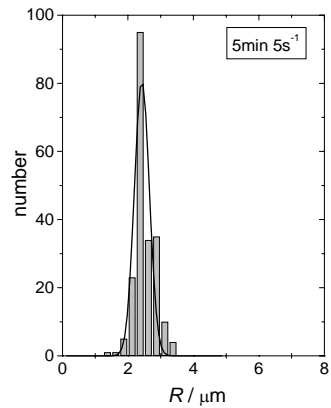
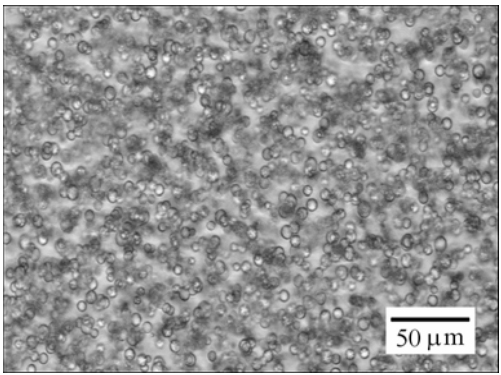
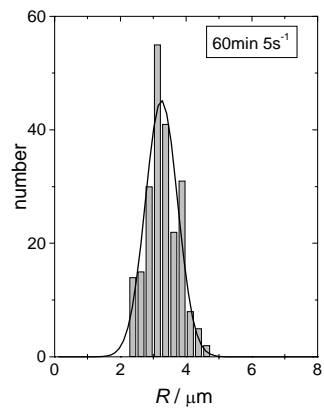

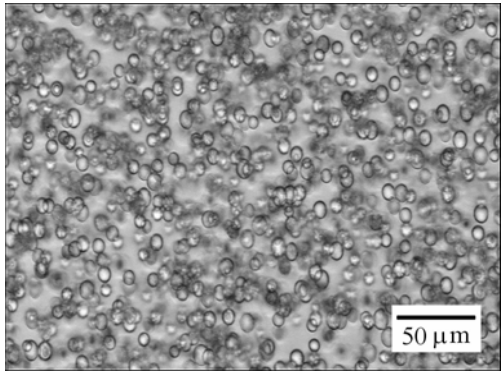 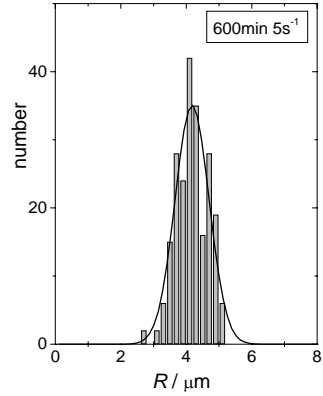

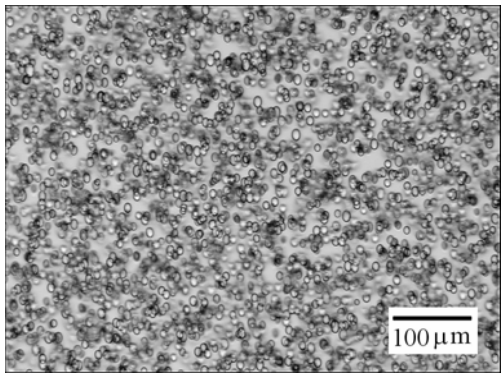 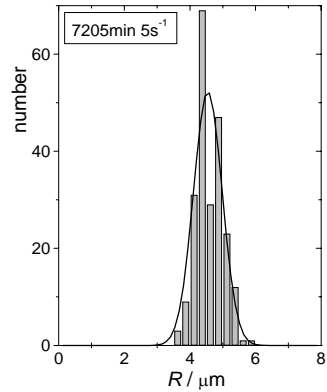

*Fig. 1: Images and drop size distributions for the system PIB 3/PDMS 152 with $\varphi_{PDMS\ 152} = 0.050$ at different times after abruptly decreasing the shear rate from 50 to 5 s$^{-1}$. The curves are Gaussian fits to the observed size distributions.*

The drop size distribution is always Gaussian for the present experiments with the following exception: In the case of low shear rates two peaks appear in the drop size evolution shortly after the step down to $\dot{\gamma}$. Typical examples are given in Fig. 2 for the blend of **COP 26\* / PDMS 48** ($\varphi_{PDMS\ 48} = 0.051$) and a reduction of $\dot{\gamma}$ from 100 to 0.5 s$^{-1}$ in terms of images and drop size distributions, and in Fig. 3 in terms of the Gaussian fits of the drop size distributions for $\dot{\gamma} = 2$ s$^{-1}$. In the latter case one observes two peaks after 2 min, one at 1.7 µm and the other at 4.4 µm. The size of the smaller drops is within experimental error identical with that measured immediately after the end of pre-shearing (2.15 µm) and varies only between 1.6 to 2.7 µm irrespectively of shear rate and the amount of drop phase ($\varphi_{PDMS\ 48} = 0.051$ or 0.150). Furthermore the position of the peak turns out to be independent of time but its height decreases monotonously until the small drops finally disappear as can be seen in Fig. 3. The second peak at higher $R$ moves towards larger radii as time advances; in doing so its height decreases and the width at half maximum becomes broader.

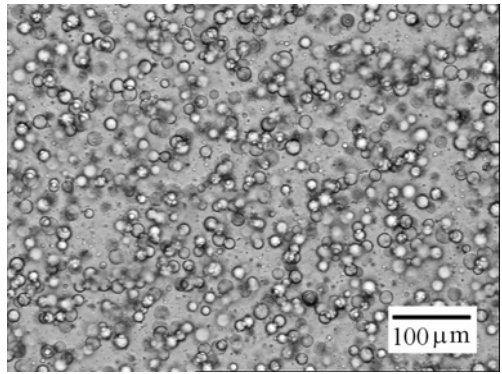 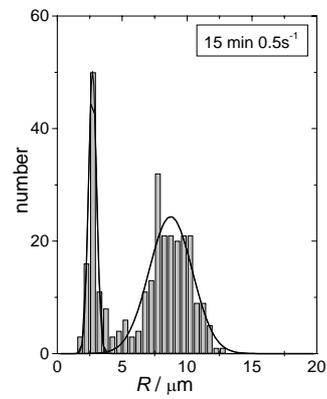

15 min

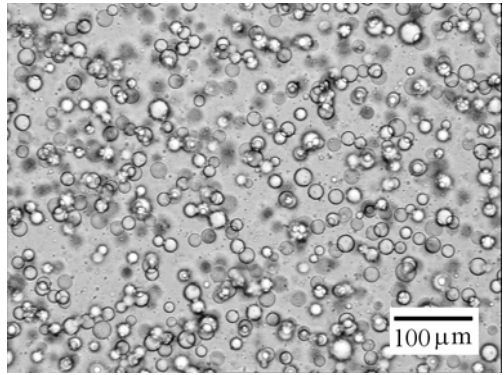 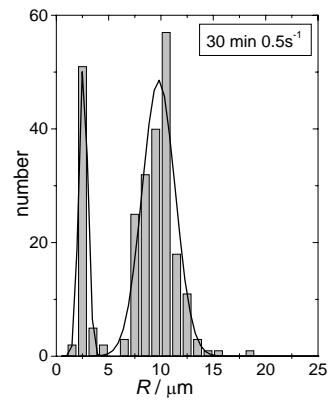

30 min

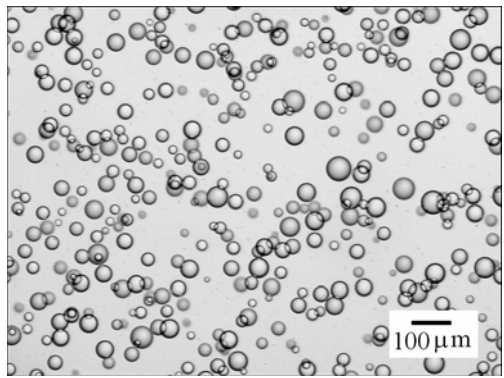 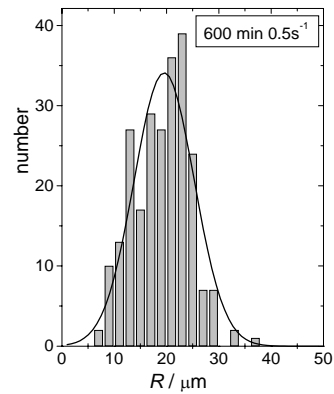

600 min

*Fig. 2: Light microscopic images and drop size distribution for the system COP 26\*/PDMS 48 with $\varphi_{PDMS\,48}$ = 0.051 at different times after decreasing the shear rate from 100 to 0.5 $s^{-1}$. The curves are Gaussian fits to the observed size distributions.*

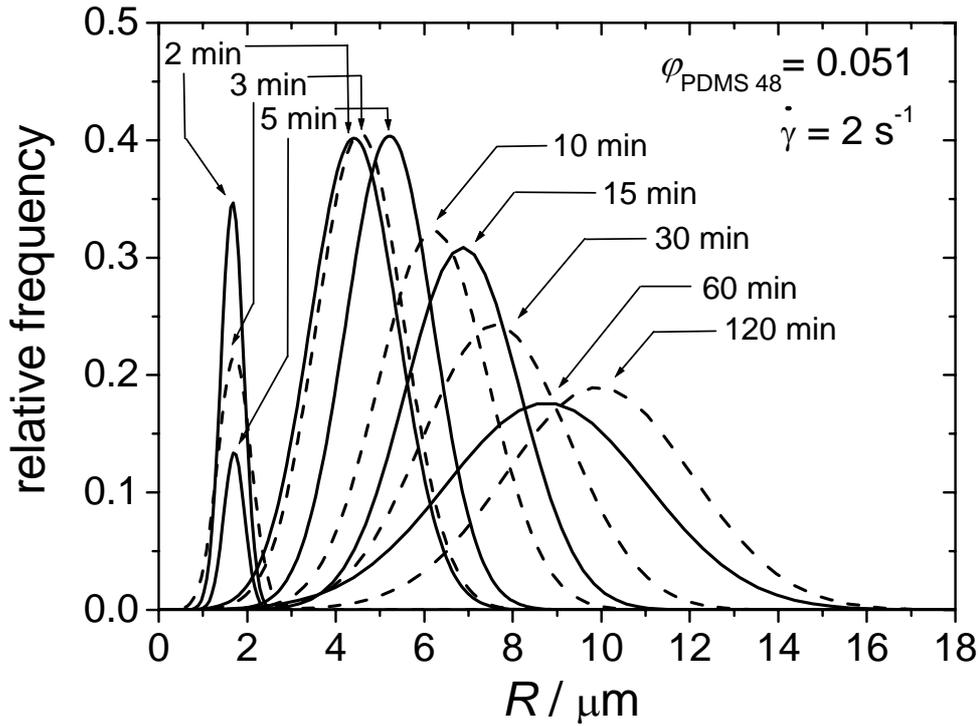

*Fig. 3: Time evolution of the Gaussian drop distribution for the system COP 26\*/PDMS 48 and $\varphi_{PDMS\,48} = 0.051$ after a step down of the shear rate from 100 to 2 s$^{-1}$.*

Experiments on the opposite side of the composition ($\varphi_{PDMS\,48} = 0.850$ and $0.949$) also reveal the occurrence of bimodal drop size distributions in the early stage of coalescence, too. Again the radii of the drops are within error equal to the radius measured directly after the end of pre-shear (1.8 μm). The conditions which lead to bimodal drop size distributions in the case of the system COP 26\*/PDMS 48 are specified in the following table.

**Table 2:** Volume fractions of COP 26\*/PDMS 48 blends and shear rates where bimodal drop distributions are observed (+)

| $\varphi_{PDMS\,48}$ | 0.049 | 0.150 | 0.850 | 0.951 |
|---|---|---|---|---|
| 10s$^{-1}$ | - | - | - | - |
| 5s$^{-1}$ | - | - | - | (+) |
| 2s$^{-1}$ | + | - | - | + |
| 1s$^{-1}$ | + | + | + | + |
| 0.5s$^{-1}$ | + | + | | |

For **PIB 3/PDMS 152** bimodality of the drop size distribution is only observed at 1 s$^{-1}$, the lowest investigated shear rate. The mean value of the peak at the smaller radii is 1.4 μm in comparison to 1.3 μm, the average drop size after pre-shear ($\dot{\gamma} = 50$ s$^{-1}$) for $\varphi_{PDMS\,152} = 0.010$. For the inverse composition this behavior is also only observed for 1 s$^{-1}$, the mean size of the smaller drops is 1.5 μm; this equals the mean radius for $\varphi_{PDMS\,152} = 0.990$ after pre-shear (1.4 μm).

The occurrence of the small peak at the radius of the morphology prevailing during the pre-shear period can be explained by the very fast growth of the size of the drops at low shear rates. This feature implies that some of the small drops do not find a partner for coalescence during that stage. Due to the fact that the coalescence efficiency decreases with increasing difference in the size of the coalescing drops [16,25] some of the drops generated during the pre-shear remain stable for long times. The present findings are in agreement with qualitative reports by Grizzuti and Bifulco [10] and Rusu and Peuvrel-Disdier [14].

*4.2. Time evolution of the mean drop size*

Fig. 4a shows how the average drop sizes change with time for the system COP 26*/PDMS 48 and $\varphi_{PDMS\ 48} = 0.051$ at different constant shear rates after a step down from 100 s$^{-1}$; Fig. 4b displays this dependence as a function of strain $\gamma = \dot{\gamma}t$.

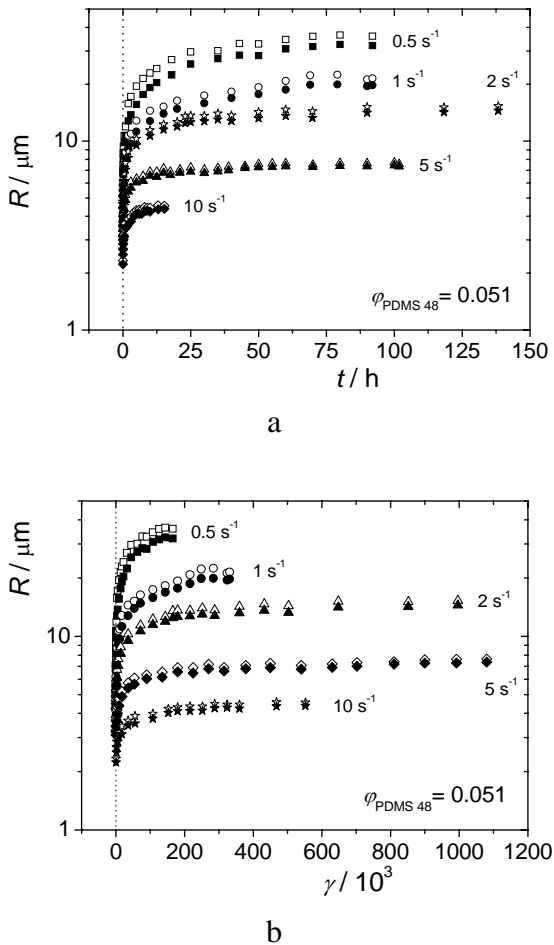

Fig. 4:  $R_V$, $R_N$ and $R_G$ for the system COP 26*/PDMS 48 and $\varphi_{PDMS\ 48} = 0.051$ as a function of time (part a) and of strain (part b) after step down of the shear rate from 100 s$^{-1}$ to the indicated shear rates. Open symbols label $R_V$, solid ones $R_N$.

In most cases it needs about 200 000 – 300 000 strain units to reach almost constant drop sizes. This range compares as follows with literature reports: 100 - 100 000 by Ramic et al. [15] (for mixtures of PDMS with polypropyleneglycole (PPG), Polyethyleneglycole (PEO) or PIB and for the system PEO/PPO), 15 000 – 20 000 by Minale et al. [12] (for PIB/PDMS) and > 165 000 by Rusu and Peuvrel-Disdier [14] (for PIB/PDMS).

The dead end radii are larger for low than for high shear rates. The polydispersity of the drop size, e.g. the ratio of $R_V$ and $R_N$, is shown in Fig. 4 as a function of strain for different shear rates. It remains below 1.3 in all cases under investigation and passes a maximum for low $\dot{\gamma}$ values.

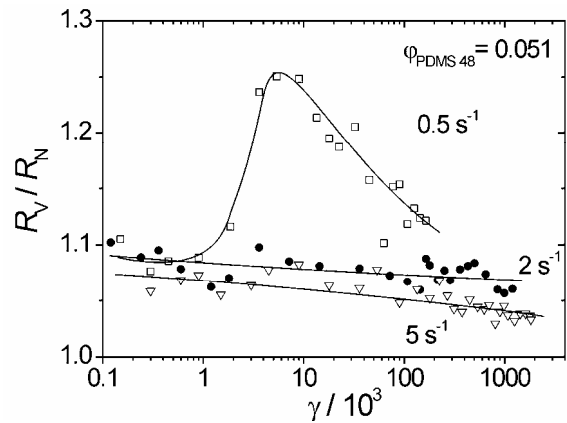

*Fig. 5:  Polydispersity of drop radii for the system COP 26*/PDMS 48 and $\varphi_{PDMS\ 48} = 0.051$ in dependence of strain after step down of the shear rate from 100 s$^{-1}$ to the indicated shear rates.*

The drop size evolution with strain for PIB 3/PDMS 152 after step down to different shear rates is shown in Fig. 6a together with the polydispersity (Fig. 6b). It takes at least 200 000 strain units to reach a constant value. Except for $\dot{\gamma} = 1$ s$^{-1}$ the polydispersity of the drop size decreases with strain. For the measurement at $\dot{\gamma} = 1$ s$^{-1}$ it is probable that the gap was not adequate.

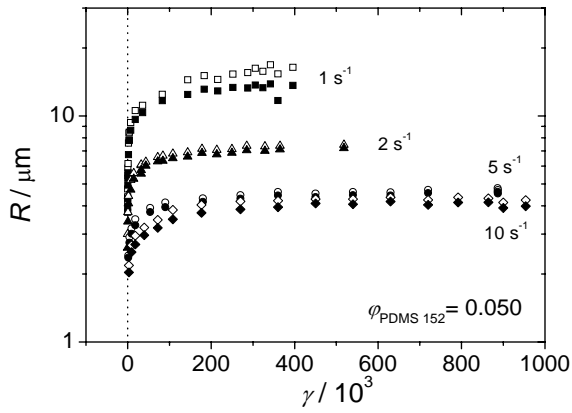

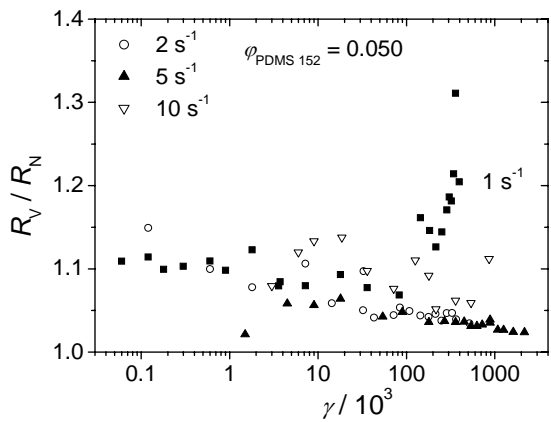

*Fig. 6: Evolution of the different drop radii $R_v$ and $R_N$ after step down of $\dot{\gamma}$ to 10, 5, 2 and 1 s$^{-1}$ (a) and $R_v/R_N$ (b) for the system PIB 3/PDMS 152 with $\varphi_{PDMS\ 152} = 0.050$ as a function of strain. In (a) the open symbols denote $R_v$ and the solid ones $R_N$.*

radii by means of eqs (5)-(7). The two curves indicate the limiting drop size for both processes: Breakup is possible for all drops with radii above the breakup curve; smaller drops remain unchanged. Coalescence, on the other hand, can take place for all drops with radii below the coalescence curve; larger drops can no longer grow. Because the slopes of these two dependences differ, the two curves intersect. This means that the diagram is divided into 4 regions: One above and one below both curves. Here only one process can occur at a time: either breakup or coalescence. The third region lies below the breakup but above the coalescence curve. Here nothing will happen to the drop because it is too big for coalescence and also too small for breakup. In the forth region, finally, both processes are physically possible.

### 4.3. Elmendorp diagrams

Fig. 7 and Fig. 8 present the results according to Elmendorp [26]. Such graphs depict the breakup curve calculated via the equation of DeBruijn (eq (4)) and eq (1) together with the coalescence curves, which are obtained by fitting the constant drop

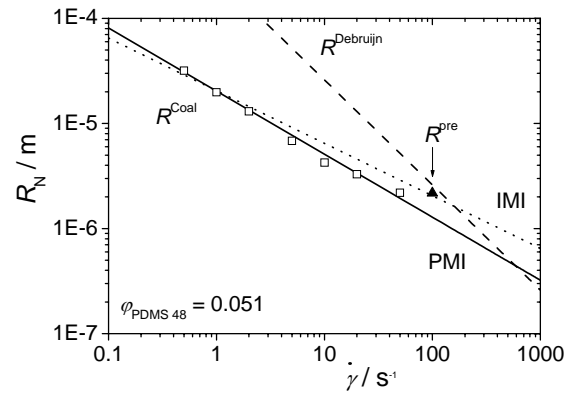

*Fig. 7: Stationary $R_N$ in dependence of the shear rate for the system COP 26*/PDMS 48 and $\varphi_{PDMS\ 48} = 0.051$; the dashed line is the breakup curve according to DeBruijn, the dotted and the solid lines are the coalescence curves according to the IMI- and the PMI-model with $h_c = 1.087$ and $0.126\ \mu m$ respectively. The triangle indicates the drop size after pre-shear at 100 s$^{-1}$.*

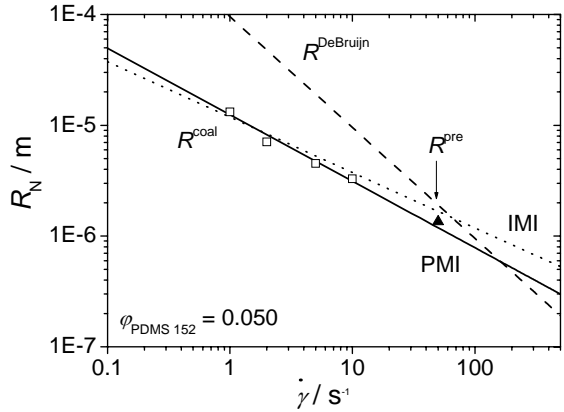

*Fig. 8: Stationary $R_N$ as a function of the shear rate for the system PIB 3/PDMS 152 and $\varphi_{PDMS\ 152}$ = 0.050; the dashed line is the breakup curve according to DeBruijn, the dotted and the solid lines are the coalescence curves according to the IMI- and the PMI-model with $h_c$ = 1.814 and 0.827 µm respectively. The triangle indicates the drop size after pre-shear at 50 s$^{-1}$.*

For both systems the coalescence radii are fitted best by the PMI-model in accordance with literature [12,14,20]. Fig. 7 and Fig. 8 show that all experiments were carried out far away from the breakup curve and therefore the final drop size is reached via coalescence only. The drop size of the respective inverse blends (i.e. 5 vol.-%PDMS 48 in COP 26* and 5 vol.-%PDMS 152 in PIB 3) are within 2 µm the same. A possible explanation for these results could be the following: The coalescence rate is determined by two parameters: the probabilities of collision and of coalescence. In the model of Janssen the collision probability is considered to be unity, which means that the critical film thickness remains as the only decisive parameter. However, according to this assumption the drop size should not depend on concentration, in contrast to the experimental observation [11,22,27,28]. For this reason one is obliged to postulate a decrease of the collision probability due to declining drop numbers. Especially at low fractions of the suspended phase the drop concentration falls so rapidly with advancing coalescence that the collision frequency decreases strongly before the size of the drop becomes the limiting factor for coalescence. For inverse blends with $\lambda$ not far away from unity the viscosity of the components is similar and the interfacial tension is (neglecting possible effect of the non-uniformity of the polymer) identical. Therefore the descent of collision frequency happens at approximately the same drop numbers per volume and consequently at similar drop size. This hypothesis can also explain the findings of Rusu and Peuvrel-Disdier [14] and Grizzuti and Bifulco [10] who reported that the drop size of the inverse blends does not differ for binary blends of PIB and PDMS at 1 and at 10 vol.-% minor phase. The present proposition is also backed by the observation that an augmentation of the drop concentration for the system COP 26*/ PDMS 48 from $\varphi_{COP\ 26*}$ = 0.051 to 0.150 results in a doubling of the drop size.

## 5. Conclusion

In this work we have studied the shear induced coalescence of the drops of the minor phase at different shear rates for one blend consisting of two homopolymers and one homopolymer/copolymer mixture. In both cases the components were practically immiscible and the molar masses of the components such that their viscosity ratios are close to unity; here we have dealt in detail with $\lambda$ > 1. For low contents of the minor phase one observes dead end drop radii in all experiments, i.e. after some time the drops do practically no longer increase in size. In our view this behavior is primarily a consequence of the low volume fractions resulting in low collision probabilities and not caused by a decrease of coalescence probability discussed in the literature. This hypothesis can also explain the observed equality of the drop sizes of inverse blends reported in literature for $\varphi_{drop}$ = 0.01 and 0.10. There is, however, no doubt that the coalescence probability can become the dominant factor at higher concentrations of the drop phase.

In the early stages of coalescence the drop size distribution becomes bimodal for low shear rates and volume fractions of the suspended phase. This observation can be explained by the coaction of two factors: (i) the increase in growth rate with a diminution of shear rate and (ii) the maximum in the coalescence probability for drops of equal size. In other words: Most of tiny drops a time zero (immediately after pre-shear) disappear rapidly because of the high coalescence probability, but some are left and for them it is time consuming to find partners, which match their size reasonably. This leads to the observed bimodality with one peak at approximately the original drop size and a main peak at large drop radii. The height of the first peak decreases continuously until it vanishes in the course of the experiment.

In the present work we could study one of the limiting cases of coalescence, namely the development of dead end drop dimensions. This was made possible by selection the viscosity ratio larger than unity, which means that the breakup curve in the Elmendorp diagram is located at comparatively large radii. This feature enables large step down intervals in shear rate and consequently large differences between the initial drop radii and the radii at which break up processes commence. There exist two possibilities for the extension of the present study to the establishment of stationary state radii. One option consist in an increase of the concentration of the drop phase (shift the coalescence curve to larger radii) another in the reduction of its viscosity (shift of the breakup curve to smaller radii).

## 6. Acknowledgment

The authors would like to thank the "Arbeitsgemeinschaft industrieller Forschungsvereinigungen", which finances this research (project numbers 12216N and 13792N).